\documentclass{elsarticle}
\usepackage{graphicx,color}
\usepackage{amsmath,amssymb}

\begin{document}
\begin{frontmatter}

\title{Comments About the Electromagnetic Field in Heavy-Ion Collisions}

\author[rbrc,bnl,ccnu]{L. McLerran } 
\author[bnl]{V. Skokov} 

\address[rbrc]{RIKEN BNL, Brookhaven National Laboratory, 
Upton, NY 11973}
\address[bnl]{Physics Department, Brookhaven National Laboratory,
	Upton, NY 11973, USA}
\address[ccnu]{Physics Department, China Central Normal University, Wuhan, China}

 \date{\today}

\begin{abstract}
In this short article we discuss the properties of  electromagnetic fields 
in heavy-ion collisions and consequences for observables. We address 
quantitatively the issue of the magnetic field lifetime in a collision 
including the electric and chiral magnetic conductivities. We show that for 
reasonable parameters, the magnetic field created by spectators in a collision
is not modified by the presence of matter. Based on this, we draw definite conclusions on observed effects in 
heavy-ion collisions. 
\end{abstract}


\end{frontmatter}

\section{Introduction}
The experiments with heavy-ion collisions of ultra-relativistic energies 
probe not only matter in extreme temperatures and densities, but also under 
action of extremely strong electro-magnetic  fields with magnitude of the hadronic scale,
$ e B \sim m_\pi^2 $~\cite{Kharzeev:2007jp,Skokov:2009qp}.

The magnetic field is a key ingredient for many observables related to  
local parity and charge parity violation~\cite{Kharzeev:2007jp}. 
The lifetime of the magnetic field, which is needed to describe the 
observed data of the elliptic flow dependence for positive and negative charged particles 
on the asymmetry, within the framework of the Chiral Magnetic Wave  must be as large as 4 fm/$c$
~\cite{Burnier:2012ae}. However, the photon azimuthal anisotropy measured at the top RHIC energy 
can be described by the magnetic field at a time scale of a few $0.1$ fm/$c$ \cite{Basar:2012bp}. 
This apparent discrepancy demands  theoretical studies of the time dependence of the magnetic field. In this short article 
we will consider effects of finite conductivity on the lifetime. We also discuss the 
dependence of the magnetic field on the collision energy and draw some conclusion on photon azimuthal anisotropy 
for  RHIC and LHC energies.

\section{ The Electro-Magnetic field in heavy-ion collision}

The Maxwell equation describing the time evolution of the electromagnetic field in a collision reads
\begin{eqnarray}
&&\frac{\partial \vec{B}}{\partial t} = - \vec{\nabla} \times \vec{E} ,\\
&&\frac{\partial \vec{E}}{\partial t} = \vec{\nabla} \times \vec{B} - \vec{j}; 	  
\label{Maxwell}
\end{eqnarray}
where the electromagnetic current can be decomposed in two pieces:
\begin{equation}
\vec{j} = \vec{j}_{\rm ext} +  \vec{j}_{\rm int},  
\label{j}
\end{equation}
the internal current, $\vec{j}_{\rm int}$, 
and the external one of the colliding nuclei, $\vec{j}_{\rm ext}$. The latter we will treat in the eikonal approximation    
neglecting effects of the proton deceleration and/or stopping. 
For  later convenience let us separate the electromagnetic field also in the ``external'' and ``internal'', e.g.
\begin{equation}
E = E_{\rm ext} + E_{\rm int}, 
\label{Esep}
\end{equation}
where the external electromagnetic field satisfies the equations
\begin{eqnarray}
&&\frac{\partial \vec{B}_{\rm ext}}{\partial t} = - \vec{\nabla} \times \vec{E}_{\rm ext} ,\\
&&\frac{\partial \vec{E}_{\rm ext}}{\partial t} = \vec{\nabla} \times \vec{B}_{\rm ext}- \vec{j}_{\rm ext}; 	  
\label{MaxwellExt}
\end{eqnarray}
while for  the internal we get 
\begin{eqnarray}
&&\frac{\partial \vec{B}_{\rm int}}{\partial t} = - \vec{\nabla} \times \vec{E}_{\rm int} ,\\
&&\frac{\partial \vec{E}_{\rm int}}{\partial t} = \vec{\nabla} \times \vec{B}_{\rm int}- \vec{j}_{\rm int}; 
\label{MaxwellInt}
\end{eqnarray}
This representation is especially convenient for finding a numerical solution due to the following two reasons. 
First, there is no need in solving the first couple of equations for the external components with the singular source terms, 
since the solution can be obtained by boosting the electric field of both nucleus. Second, the singularities of the 
sources are spread by the fields, which leads to a better convergence of a numerical scheme.

In the matter rest frame, the internal current may include the contribution from the Ohmic conductivity 
\begin{equation}
\vec{j}_{\rm Ohm} = \sigma \vec{E} 
\label{Ohm}
\end{equation}
and the one induced by the QED anomaly 
\begin{equation}
\vec{j}_{\rm anom} = \sigma_{\chi} \vec{B}, 
\label{anom}
\end{equation}
where $\sigma_{\chi} \propto  \mu_5$ and $\mu_5$ is the axial chemical potential, that can be created by
a sphaleron  transition in a collision. The value of $\mu_A$ is quite uncertain, so is $\sigma_\chi$. The electric 
conductivity $\sigma_{\rm Ohm}$ was calculated using a first principal lattice QCD approach in the {\it quenched} approximation. 
It has been found~\cite{Ding:2010ga} that above the transition temperature $T_c$\footnote{Similar result was obtained in Ref.~\cite{Aarts:2007wj} earlier.}:
\begin{equation}
\sigma^{\rm LQCD}_{\rm Ohm} = (5.8\pm2.9) \frac{T}{T_c} {\rm MeV}. 
\label{sigma_qcd}
\end{equation}
The quenched approximation used in Ref.~\cite{Ding:2010ga} does not allow to draw a final concussion on the magnitude of the 
conductivity. We, however, do not expect that taking into account the quark loops will modify the results larger then by a 
factor of  2.

We also assume that charged matter, if created at early times, does not  develop large collective velocity $v$, which allows us to neglect the contribution $\vec{v}\times\vec{B}$ when transforming to the laboratory rest frame. 
Both conductivity $\sigma_{\rm Ohm}$ and $\sigma_{\chi}$ may result in a substantial increase of the lifetime
of the magnetic field.  In what follows we consider both cases separately.

{\it Magneto-hydro scenario}, $\sigma_\chi=0$ and $\sigma_{\rm Ohm}\gg 1/t_c$, where $t_c$ is the characteristic timescales. For this case,  
the Maxwell equations can be written as 
\begin{equation}
\nabla^2 \vec{B}_{\rm int}= \frac{\partial^2 \vec{B}_{\rm int}}{\partial t^2} + \sigma_{\rm Ohm} \frac{\partial \vec{B}_{\rm int}}{\partial t} + \sigma_{\rm Ohm}\frac{\partial \vec{B}_{\rm ext} }{\partial t}. 
\label{d2B}
\end{equation}
For late times the external contribution, the last term, is not important and can be skipped.  
The first term on the right-hand side of Eq.~\eqref{d2B} can be neglected for  $\sigma_{\rm Ohm}\gg 1/t_c$  
yielding a diffusion equation for the components of the internal magnetic field. 
The diffusion equation leads to a significant increase of the lifetime of the magnetic field, as was pointed out  
in Ref.~\cite{Tuchin:2013ie} recently. However, this result was obtained under the assumption of 	  $\sigma_{\rm Ohm}\gg 1/t_c$. 
The characteristic time scale is defined by the external magnetic field and proportional to the sickness of the nucleus in the beam direction, i.e. $t_c \sim 2R/\gamma$. For the top RHIC energy, $t_c\sim 1/5$ fm $= 10^{-3}$ MeV$^{-1}$. The condition    
$\sigma_{\rm Ohm}\gg 1/t_c \sim 10^3$ MeV, therefore can be rewritten as a condition on the temperature of the created plasma, 
$T/T_c\gg10^2$, which is unrealistic for collisions of such energy. Thus we conclude that for realistic values of the parameters 
the condition $\sigma_{\rm Ohm}\gg 1/t_c$ is  never satisfied. In the next section, we will demonstrate our numerical results for realistic values of  $\sigma_{\rm Ohm}$, illustrating the same fact.   
\begin{figure}
\centerline{\includegraphics[scale=0.5]{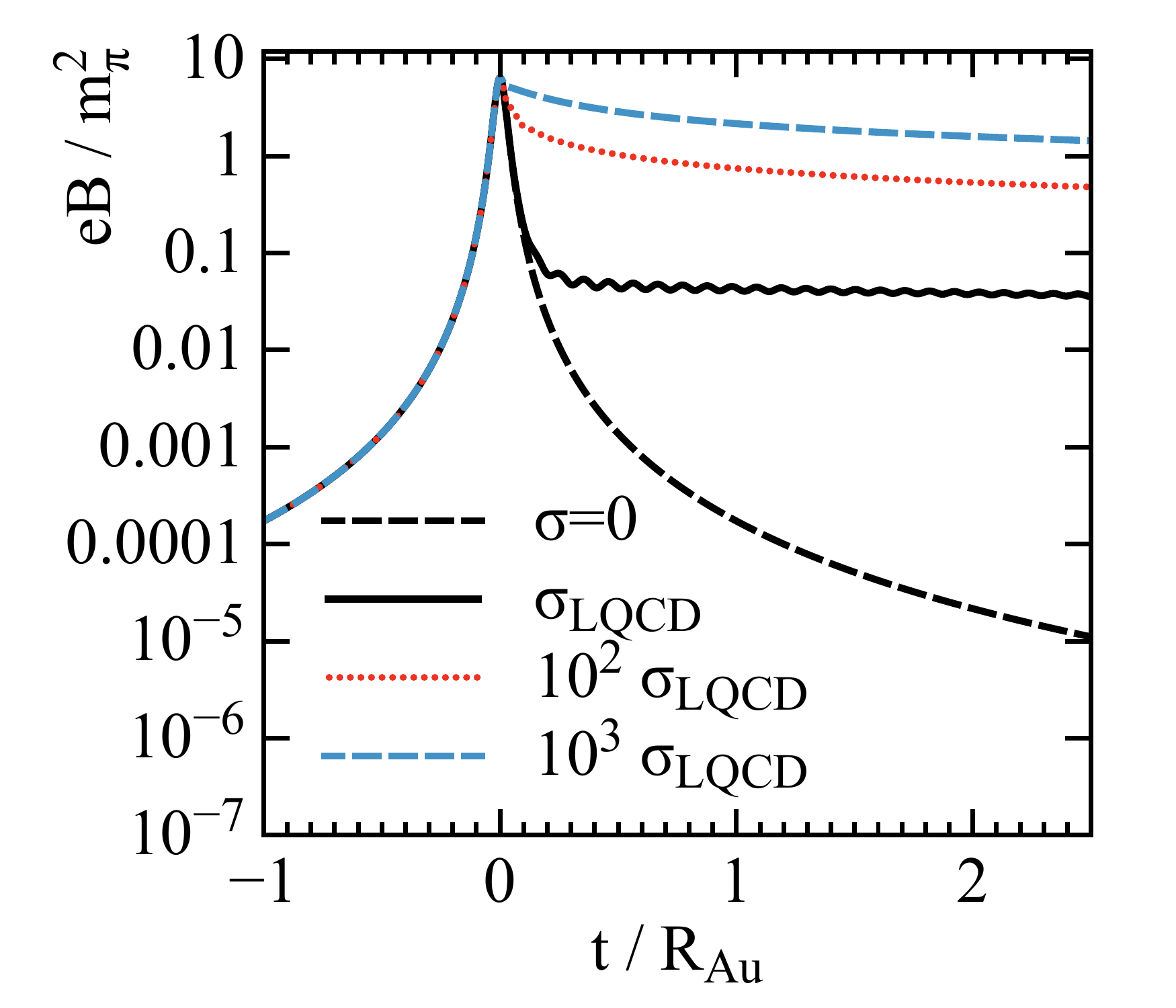}}
\caption{
Magnetic field for static medium with Ohmic conductivity, $\sigma_{\rm Ohm}$.
}
\label{fig_Magnetic_field_ohmic}
\end{figure}

{\it Formation of knots},  $\sigma_\chi\gg 1/t_c$ and $\sigma_{\rm Ohm}=0$. In this case, neglecting second
derivatives with respect to time we get    
\begin{equation}
\nabla^2 \vec{B}_{\rm int} + \sigma_\chi \nabla \times  \vec{B}_{\rm int}  + \sigma_\chi \nabla \times  \vec{B}_{\rm ext}=0  
\label{Bknot}
\end{equation}
As in previous case, for late times, the external contribution can be neglected resulting in
\begin{equation}
 \vec{B}_{\rm int} = - \frac{1}{\sigma_\chi}  \vec{\nabla} \times \vec{B}_{\rm int}. 
\label{B_knot_froz}
\end{equation}
The total magnetic helicity  
\begin{equation}
H = \int_V d^3 x \vec{A}\  \vec{B}
\label{H}
\end{equation}
is conserved for the closed systems. 
For two fluxes of the magnetic field $\phi_1$ and $\phi_2$,
the helicity can be related to the linking number $H=2 n \phi_1 \phi_2$.
Substituting Eq.~\eqref{B_knot_froz} to Eq.~\eqref{H} and performing trivial transformations 
we obtain for the helicity  
\begin{equation}
H_{\chi} = - \frac{1}{\sigma_\chi} \int_V \vec{B}^2 d^3x = - \frac{8\pi}{\sigma_\chi} E_B,  
\label{Hchi}
\end{equation}
where $E_B$ is the total magnetic energy. This shows that conservation of the helicity leads to the 
conservation of the total magnetic energy for the processes with the timescales, $t_c\ll1/\sigma_\chi$. 
The volume $V$ in Eq.~\eqref{Hchi} is defined by the region of space, where $\sigma_\chi\ne0$. 
Owing to the expansion of the medium this volume grows in time roughly as $t^3$ for late times, as $t$ for 
early times. Therefore
we expect the magnetic field to decay according to the power law $B\sim t^{-3/2}$ or $B\sim t^{-1/2}$ . This is 
somewhat slower then the decay of the field induced by the spectators $B_{\rm spect} \sim t^{-2}$.  
This conclusion, however, does not take into account the formation of non-trivial topological 
objects, knots of the magnetic field with non-trivial linking number. As was shown in Ref. ~\cite{PhysRevE.84.016406}, 
the higher the linking number corresponds to longer  lifetime of the magnetic field up to $B_{n}\sim t^{-1/6}$. 
Returning to the constraint $\sigma_\chi\gg 1/t_c$, we can roughly estimate if this is satisfied in
heavy ion collisions. The chiral conductivity is defined by chiral chemical potential 
following~\cite{Kharzeev:2009pj}
\begin{equation}
\sigma_\chi = \left(\frac{e^2}{2\pi^2} N_c \sum_f q_f^2 \right) \mu_5. 
\label{sigma_chi}
\end{equation}
To make an optimistic estimate we will consider quite high values of $\mu_5\approx 1$ GeV.
Nevertheless even in this case the numerical value for $\sigma_\chi$ is only 
$15$ MeV, while the inverse characteristic time for a collision of the top RHIC energy 
is $10^3$ MeV. This again shows that the effects of finite  $\sigma_\chi$ will not be important for 
the top RHIC and LHC energies. 

\section{Numerical results }
Numerical results to be presented in this section are obtained with the Yee algorithm~\cite{1138693}, which 
is numerically stable for the conductivity ranging from 0 to $\infty$. The calculations proceed as follows. 
First,  we initialize the distribution of protons both in projectile and target according to 
the Woods-Saxon distribution with the standard parameters~\cite{Alver:2008aq}. We also 
checked that up to statistical fluctuations the results obtained using the Wood-Saxon distribution 
coincide with those in approximation of collision of homogeneously charged spheres, which 
will be used for the sake of illustration in Fig. 1.
Next we assume that both
target and projectile move with the opposite velocities of the same magnitude $v^2=1-(2m_p/\sqrt{s})^2$. 
The conducting medium in the collision is not formed immediately, because the quarks need time to be created 
from the glasma field. Nonetheless, to make our estimates of the conductivity effects as optimistic as possible 
we will consider that the conducting medium is formed immediately after the collision and does not alter 
during the evolution. We also neglect 
possible non-equilibrium effects and finite time response of the medium to the electric field, which 
as was shown in Ref.~\cite{Cassing:2013iz} may play an important role.

The decay of the conductivity owing to expansion of the medium can only decrease the 
lifetime of the magnetic field and thus will not be considered here. 
Our simulations are done for Au-Au collisions at energy $\sqrt{s}=200$ GeV and fixed impact parameter $b=6$ fm. 
In Fig. 1 we show time evolution of the magnetic field in the origin $\vec{x}=0$ as a function of the electric 
conductivity $\sigma_{\rm Ohm}$. The results show that the lifetime of the strong magnetic field ($eB>m_\pi^2$) is not affected 
by the conductivity, if one uses realistic values obtained in Ref.~\cite{Ding:2010ga}.

\begin{figure}[t]
\centerline{\includegraphics[scale=0.5]{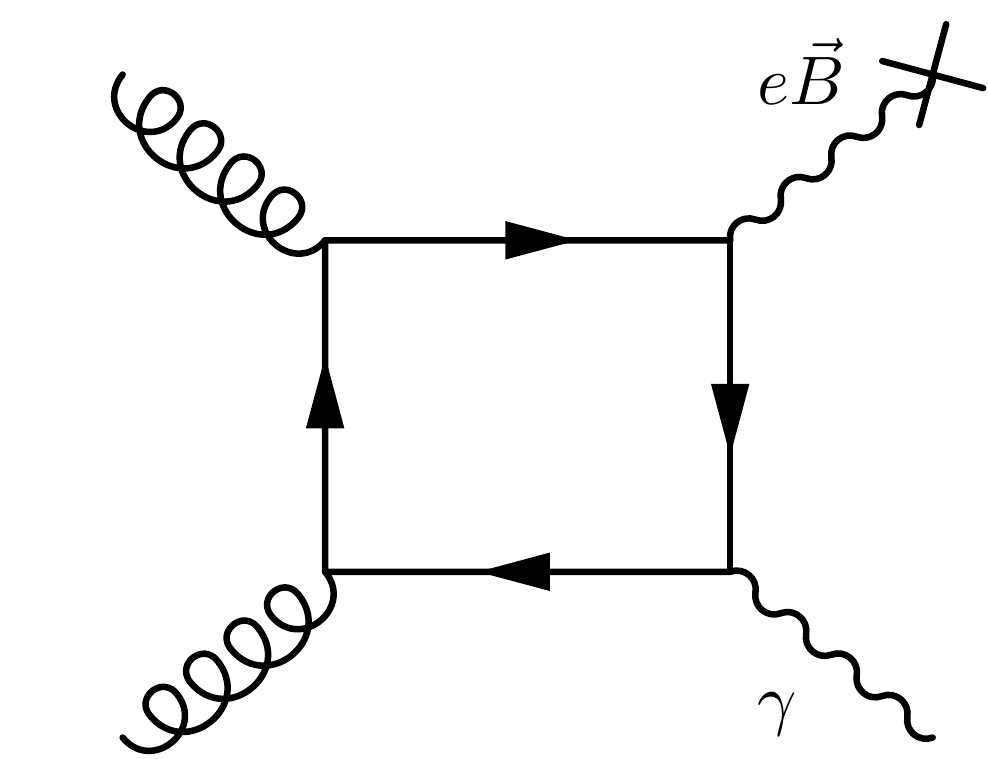}}
\caption{
Coupling of gluon field to photons. The cross labels external coherent magnetic field line.  
}
\label{fig_box}
\end{figure}
\section{Energy dependence}
In the previous section, we established that for realistic values of the conductivities the electromagnetic fields 
in heavy-ion collisions are almost unmodified by the presence of the medium. Thus one can safely use the magnetic 
field generated by the original protons only. This magnetic field can be approximated as follows
\begin{equation}
e B(t, \vec{x}=0) = \frac{1}{\gamma}\frac{c Z}{ t^2 + (2 R/\gamma)^2},  
\label{eb}
\end{equation}
where $Z$ is the number of protons, $R$ is the radius of the nuclei, $\gamma$ is the Lorentz factor and, finally,  $c$ is some non-important numerical coefficient.
We are interested on the effect of the magnetic field on the 
matter. Thus we need to compute the magnetic field at the time
$t_m$, characterising matter formation time. On the basis of a  very general argument, 
one would expect that $t_m= a Q_s^{-1}$. The phenomenological constraints from photon azimuthal anisotropy at the top RHIC energy
demand $t_m\approx 2R/\gamma_{\rm RHIC}$, i.e. $a=2R Q^{\rm RHIC}_s/\gamma_{\rm RHIC}$. Using this relation, we can estimate 
the magnitude of the magnetic field at the LHC energies at the time $t=t_m$. For the LHC, since $t_m\ll 2R/\gamma$ we have 
\begin{equation}
eB_{\rm LHC}/eB_{\rm RHIC}  = 2 \frac{\gamma_{\rm RHIC}}{\gamma_{\rm LHC}} \left( \frac {Q_s^{\rm LHC}}{ Q_s^{\rm RHIC} } \right)^2. 
\label{LHC}
\end{equation}
Using $Q_s\propto (\sqrt{s})^{0.3}$, we get $(eB_{\rm LHC}/eB_{\rm RHIC})^2\approx0.32$.
Therefore the magnetic field is not suppressed strongly at matter formation time at LHC. 
However, the anisotropy of photon production from magnetic field is suppressed. 
This follows from the following 
dimensional argument.
First of all, to couple gluons to photons we are considering the box diagram showed in 
Fig.~\ref{fig_box} (see  also Ref.~\cite{Basar:2012bp}). 
Integrating quark fields, we obtain an effective coupling $ g_s^2 e^2  h  G^2 F^2 $, where 
$F^2=F_{\mu\nu} F^{\mu\nu}$ and similar for $G^2$. At early times, according to saturation picture 
the contribution, $g_s^2 G^2$ is of the order 1 in powers of $g_s$. One of the factors of $F_{\mu\nu}$ is associated with 
external coherent magnetic field, thus it compensates one power of $e$. Therefore, the diagram 
is of the order $g_s^0 e$, i.e. of the same as Coulomb or Compton contributions to photon production.   
The prefactor $h$ is defined by the momentum running in the loop. In the first approximation it is proportional
to $Q_s^{-4}$ and canceled out by the gluon $G^2$. Finally, parametrically we have that the amplitude of photon production 
is $e B k_\perp$ (we consider only central rapidity region, $k_z=0$). The rate of photon production is thus 
$(e B)^2 k_\perp^2$. 

Thus the photon $v_2$ generated at early stage is thus, $B^2 k_\perp^2/(Q_s^6)$. Therefore, the ratio 
\begin{equation}
v_2^{\rm LHC}/v_2^{\rm RHIC} =  \left( \frac {Q_s^{\rm RHIC}}{ Q_s^{\rm LHC} } \right)^6  \left( \frac{eB_{\rm LHC}}{eB_{\rm RHIC} }\right)^2
\label{v2}
\end{equation}
calculated at given $k_\perp$ is small owing to the prefactor  $\left(\frac {Q_s^{\rm RHIC}}{ Q_s^{\rm LHC} }\right)^6$.

\section{Conclusions}
In this article, we calculated the time dependence of the magnetic field on time allowing for conductivity effects 
in the plasma. In contrast to the results obtained in Ref.~\cite{Tuchin:2013ie}, we showed that 
the effects of conductivity do not play an important role for realistic values. We also have argued that while photon flow may 
receive reasonable size effects from production in an external magnetic field at RHIC energies, the induced flow is very small at
LHC energies.
This can also be tested experimentally with the method proposed in Ref.~\cite{Bzdak:2012fr}.

\section{Acknowledgments}
We thank Adam Bzdak, Dmitri Kharzeev, Jinfeng Liao and Shu Lin for stimulating discussions.
The research of the authors  is supported
by the U.S. Department of Energy under contract \#DE-AC02-98CH10886.


\end{document}